\documentstyle[aps,prl,epsf]{revtex}
\begin{document}
\twocolumn[\hsize\textwidth\columnwidth\hsize\csname@twocolumnfalse\endcsname
\title{Coherent Low-Energy Charge Transport in a Diffusive S-N-S Junction}
\author{P. Dubos, H. Courtois, O. Buisson and B. Pannetier}
\address{Centre de Recherches sur les Tr\`es Basses Temp\'eratures,
C.N.R.S., associ\'e \`a l'Universit\'e Joseph Fourier, \\ 
25 Av. des Martyrs, 38042 Grenoble, France}
\date{\today}
\maketitle
\begin{abstract}
 We have studied the current voltage characteristics of diffusive
 mesoscopic Nb-Cu-Nb Josephson junctions with highly-transparent Nb-Cu
 interfaces.  We consider the low-voltage and high-temperature regime
 $eV<\epsilon_{c}<k_{B}T$ where $\epsilon_{c}$ is the
 Thouless energy.  The observed excess current as well as the observed
 sub-harmonic Shapiro steps under microwave irradiation suggest the
 occurrence of low-energy coherent Multiple Andreev Reflection (MAR).
\end{abstract}]

A S-N-S junction made of a normal metal (N) embedded between two
superconducting electrodes (S) exhibits a zero-voltage Josephson
supercurrent \cite{Likharev-rsj}.  At the N-S interface, the
microscopic mechanism is the Andreev reflection of an electron into a
hole which traces back almost exactly the trajectory of the incident
electron.  This coherent process corresponds to the transfer of a
Cooper pair in S and its inverse to the diffusion of an electron
Andreev pair in the normal metal.  Let us consider a mesoscopic
diffusive normal metal where the length $L$ is larger than the elastic
mean free path $l_{e}$ but smaller than the phase-coherence length
$L_{\varphi}$.  In the long junction regime $L \gg \sqrt{\hbar
D/\Delta}$ of interest here, the relevant energy scale is the Thouless
energy $\epsilon_{c}= \hbar / \tau_{D} = \hbar D/ L^2 \ll \Delta$. 
Here $D=v_{F}l_{e}/3$ is the diffusion coefficient, $\tau_{D} =L^2/D$
is the diffusion time, $\Delta$ is the superconducting gap.  The
essential fact is that Andreev pairs with a low energy obeying
$\epsilon<\epsilon_{c}$ remain coherent over the whole normal metal
length $L$ \cite{Courtois96}.  This coherent window contributes
significantly to phase-coherent transport even if the thermal
distribution width $k_{B}T$ is much larger than the Thouless energy
$\epsilon_{c}$.  This is exemplified in Andreev interferometers by the
large magnetoresistance oscillations with an amplitude of about
$\epsilon_{c}/k_{B}T$ \cite{Courtois96}.  Such a weak power-law
temperature dependence is in striking contrast with the exponential
decay of the Josephson coupling.

At finite bias voltage ($V$), a dynamic regime takes place where the
superconducting phase is time-dependent while successive
Andreev-reflected electrons and holes gain the energy $eV$ at each
travel.  This Multiple Andreev reflections (MAR) process leads to a
sub-gap structure at the voltages $2\Delta/pe$ (p integer) in the
current-voltage characteristic \cite{BTK} and the energy distribution
function \cite{Anthore}.  In quantum point contacts, the whole process
of MAR is phase-coherent since the electron transit time is very
short.  This regime of coherent MAR was recently emphasized in the
description of both the current voltage characteristics, the shot
noise and the $d.c.$ supercurrent \cite{Goffman} of ballistic atomic
contacts \cite{Averin95,Cuevas96,Muller}.  The related non-sinusoidal
current-phase relation is indicative of coherent multiple charge
transfer, but no direct observation has been reported so far.

In a diffusive S-N-S junction, the diffusion of electrons through the
normal metal involves the diffusion time $\tau_{D}$ which must be
compared to the period of the phase difference $\chi$ evolution
$\tau_{V}=h/2eV$.  The quasi-static condition for a small phase
evolution during $n$ successive Andreev reflections at each interface
writes $2n \tau_{D} < \tau_{V}$, which is equivalent to
$eV<\pi\epsilon_{c}/2n$.  In this regime, electrons and holes in the
coherent window ($\epsilon<\epsilon_{c}$) can experience MAR while
maintaining global coherence of the induced Andreev pairs.  This
process should imply the coherent transfer of multiple charges as in
short ballistic point contacts, provided the interface transparency is
close to 1.  Recent experiments in diffusive S-N-S junctions focused
on the opposite regime $\epsilon_{c}<eV\simeq \Delta$
\cite{Strunk,Taboryski}.  Subgap structures were observed, but no
signature of coherent MAR was found.

The time evolution of the out-of-equilibrium energy distribution
induced by MAR was also considered phenomenologically, taking into
account the energy relaxation.  This lead to the prediction of an
enhancement of the junction conductance at finite voltage
\cite{ZhouSpivak,Argaman}.  The related $a.c.$ Josephson coupling at
twice the Josephson frequency was experimentally confirmed
\cite{Lehnert}.  A close analogy can be drawn with the dissipative
transport in a mesoscopic Aharonov-Bohm ring \cite{Landauer}.  Indeed
both the relaxation current in such a ring and the MAR excess current
in a S-N-S junction appear at bias voltages below $\hbar/\tau_{e}$
\cite{ZhouSpivak,Argaman,Landauer} where $\tau_{e}$ is the energy
relaxation time.

In this Letter, we present an investigation of the coherent dynamic
regime in high-quality diffusive S-N-S junctions. We focus on the
contribution of low-energy Andreev pairs to the current-voltage
characteristic.  We used as a probe an $a.c.$ microwave field with a
small photon energy $\hbar \omega<\epsilon_{c}$ so that it sits within
the coherent energy window.  Our main result is the observation of
sub-harmonic Shapiro steps that we discuss in terms of coherent
multiple pair transfer at low energy.  Here, the relevant number of
Andreev reflections $\hbar\omega/2eV$ is imposed by the photon
microwave energy, not by any energy gap.

\begin{figure}
\epsfxsize=8 cm 
\epsfbox{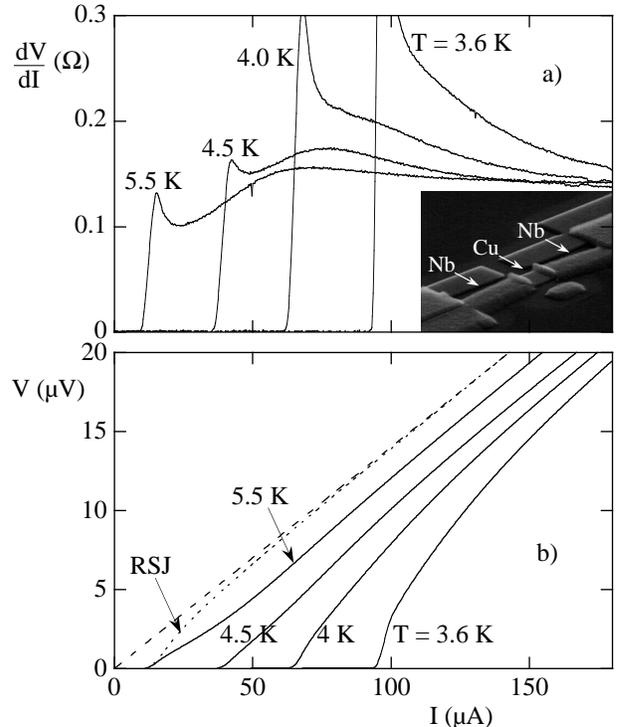}
\caption{a) Measured differential resistance of sample A as a function of
$d.c.$ bias current at different temperatures.  Inset : Electron
micrograph of a typical sample made by shadow evaporation.  b)
Current-voltage characteristics as obtained by numerical integration
of differential resistance curves.  Also shown is the theoretical
curve calculated from the RSJ model at 5.5 K (dotted line) and the
ohmic behaviour (dashed line).  The data all sit in the coherent
window $eV < \epsilon_{c}=$ 33 $\mu$eV.}
\label{RI/sample}
\end{figure}

The samples consist of a small Cu conductor attached to two Nb
electrodes which are superconducting below $T_{c} \simeq$ 8 K (Fig. 
\ref{RI/sample}a inset).  The electrical contacts include an on-chip
capacitance (0.2 pF) which connects the microwave circuit made of a
cryogenic 50 $\Omega$ coaxial cable to the sample.  The fabrication
process makes use of e-beam lithography and a shadow evaporation
technique based upon a thermostable resist compatible with UHV
electron-beam Nb evaporation \cite{DubosVictrex}.  The two samples
described here (labelled A and B) belong to the series of S-N-S
junctions investigated in Ref.  \cite{DubosIc}.  They show a similar
behavior.  We intentionally kept the thermal fluctuations negligible
at every temperature by designing wide junctions so that their
normal-state conductance $G_{N}$ and hence the Josephson energy $E_{J}
= \hbar I_{c}/2e$ is large : $E_{J} > k_{B}T_{c}$.  The sample and
measurement parameters were also chosen to fulfill the condition
$eV<\epsilon_{c}<k_{B}T<\Delta$.  The parameters for sample A
(respectively B) are the following : length of the normal metal L =
710 (800) nm, width 580 nm and thickness 100 nm.  The diffusion
coefficient is D = 250 (230) cm$^{2}$/s and the normal state
resistance is 0.152 (0.183) $\Omega$.  The calculated Thouless energy
$\epsilon_{c}$= 33 (24) $\mu$eV coincides with the values obtained
from the magnitude and temperature dependence of the $d.c.$ critical
current $I_{c}$ assuming perfect transmission at the S-N interfaces
\cite{DubosIc}.

Fig.  \ref{RI/sample}a shows the differential resistance of sample A
as a function of the $d.c.$ current bias.  The critical current
$I_{c}$ is easily identified by the sharp jump in differential
resistance, even at high temperatures when it is strongly reduced. 
The differential resistance characteristics shows a striking behavior
above 4 K : the amplitude of the main peak at the critical current
decreases while a broader bump develops at higher current.  In this
temperature range the current-voltage is fully reversible, which
discards heating effects.  The comparison in Fig.  \ref{RI/sample}b
with the ohmic behaviour and the resistively shunted junction (RSJ)
model $V=G_{N}^{-1} \sqrt{I^2 - I_{c}^2}$ (at 5.5 K) demonstrates that
this behavior corresponds to a low-energy excess current.  The
observed features are reminiscent of the foot-like structure
previously observed in superconducting micro-bridges
\cite{Tinkhambook,Lindelof} which was assigned to non-equilibrium
processes in presence of a gap oscillation.  In our long S-N-S
junctions, the role of the gap is played by the mini-gap which sets
the energy scale at $\epsilon_{c}$.  The microscopic origin of the
excess current is the coherent MAR : each Andreev reflection at the
N-S interfaces transfers a Cooper pair in and out, resulting in a
current increase until an inelastic process takes place
\cite{Volkov79}.  In average, each coherent electron transfers
$\tau_{e}/\tau_{D}$ times the elementary charge.  From the
non-equilibrium model of Ref.  \cite{Argaman}, a differential
resistance bump is predicted at $eV=\hbar/\tau_{e}$.  In the
experiment, the position of the differential resistance maximum
corresponds to $\tau_{e}$ = 90 ps at T = 5.5 K. This value is
consistent with the known electron-phonon relaxation time in Cu
\cite{Roukes} and exceeds the diffusion time $\tau_{D}$ = 20 ps at
every temperature.  It gives a phase-coherence length $L_{\varphi}
\approx$ 1.5 $\mu$m \cite{Pannetier86} which is about twice the normal
metal length.  The amplitude of the conductance enhancement $\partial
G$ was estimated by Zhou and Spivak \cite{ZhouSpivak} as : $\partial
G/G_{N} \simeq \epsilon_{c}/k_{B}T \, .  \, \tau_{e}/\tau_{D}$.  This
gives here $\partial G/G_{N} \simeq$ 0.3 at 5.5 K, in qualitative
agreement with the experiment.

\begin{figure}
\epsfxsize=8.5 cm
\epsfbox{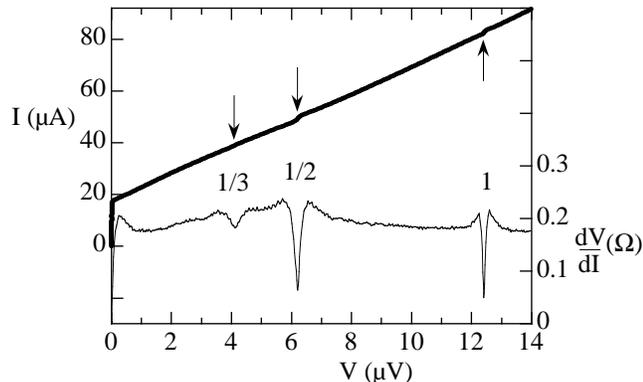}
\caption{ Lower curve (right scale) : Differential resistance of
sample B at 4 K as a function of voltage in presence of a microwave
current of frequency 6 GHz and power -20 dBm.  Structures induced by
the microwave current appear at the usual Shapiro step position $V_{1}
= \hbar \omega/2e$ = 12.4 $\mu$V and also at one half and one third of
this value.  Upper curve : the corresponding voltage steps on the
current-voltage characteristic.}
\label{microwaves}
\end{figure}

\begin{figure}
\epsfxsize=8.0 cm
\epsfbox{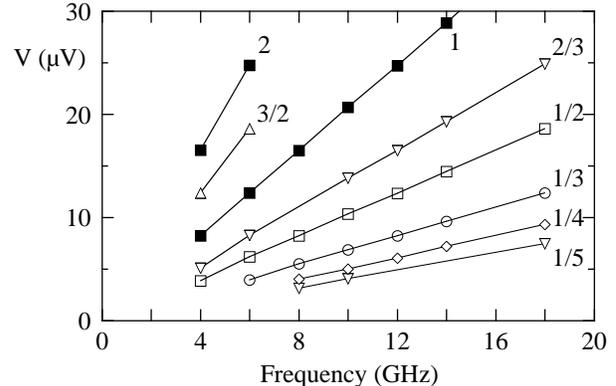}
\caption{Peak position as a function of frequency for integer (black
symbols) and non-integer steps (open symbols) in sample B. These steps
were observed at a 4K temperature and a -10 dBm microwave power.}
\label{peakposition}
\end{figure}

We investigated the coherent nature of the low voltage electrical
transport by applying a low power (-20 dBm) microwave field.  This low
power ensures us that the microwave acts only as a probe and does not
drive the electron energy distribution.  Fig.  \ref{microwaves} shows
the differential resistance and the current as a function of voltage
for sample B at 4 K in the presence of an $a.c.$ current $I_{\omega}$
of frequency 6 GHz.  The feature corresponding to the Shapiro step is
observed at voltage $V_{1} = \hbar \omega/2e$ = 12.4 $\mu$V (index 1
on the figure), as expected.  In addition, we observe two structures
at one half : $V_{1/2}$= 6.2 $\mu$V and one third : $V_{1/3}$ = 4.1
$\mu$V of this voltage (index $\frac{1}{2}$ and $\frac{1}{3}$).  The
latter is clearly visible only on the differential resistance curve. 
Because of the high temperature and the small $a.c.$ excitation, the
plateau amplitudes are small and rounded by thermal fluctuations.  The
magnitude of these structures was observed to increase linearly with
the microwave current $I_{\omega}$, leading to true plateaux with zero
differential resistance at high $a.c.$ power (-10 dBm).  Then, a large
number of peak structures, both harmonic and sub-harmonic, emerge from
the noise level.  We found that their location is independent on both
the temperature and the microwave power.  The position of the observed
peak structures as a function of the microwave frequency between 4 and
18 GHz is plotted in Fig.  \ref{peakposition}.  The harmonic 1
represents the fundamental Shapiro step.  The ensemble of the peaks
obeys the simple law given by $V= m/n \, .  \, \hbar \omega /2e$. 
While the integer steps (n = 1) are simple results of the ordinary
Josephson coupling, the observation of fractional steps (n $>$ 1)
deserves a detailed discussion.

\begin{figure}
\epsfxsize=8.0 cm
\epsfbox{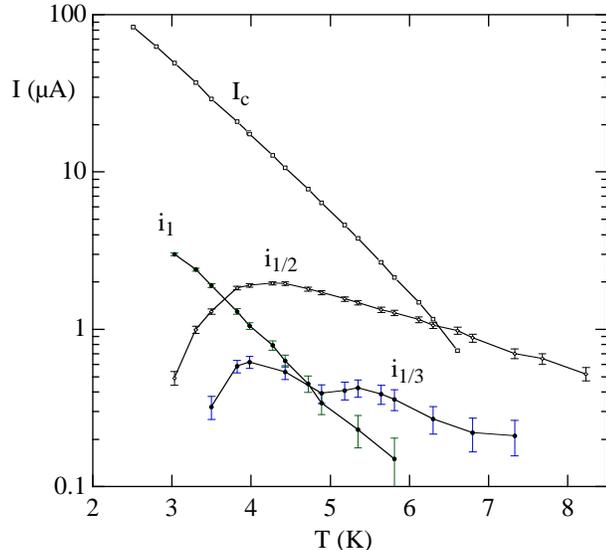}
\caption{Temperature dependence of the half width $i_{m/n}$ of integer
and fractional steps in sample B under -20 dBm microwave irradiation
at 6 GHz.  The detection threshold was about 0.1 $\mu$A. On the same
graph is plotted the measured $d.c.$ critical current $I_{c}$.}
\label{Iceff}
\end{figure}

In order to extract the step amplitudes, we assume that the steps can
be described by the RSJ model : the step width is twice an effective
critical current $i_{m/n}$ smoothed by a thermal broadening.  We used
either an analytical expression or the high temperature approximation
\cite{Likharev-rsj} to determine for each temperature and each step
the actual half width $i_{m/n}$.  Fig.  \ref{Iceff} summarizes its
temperature dependence for the three main Shapiro steps (1,
$\frac{1}{2}$ and $\frac{1}{3}$) at 6 GHz microwave frequency.  At the
center of the step 1 plateau, the current bias $I \simeq G_{N} V_{1}$ is much
larger than the critical current $I_{c}$, so that the step amplitude
should follow the usual voltage-bias law $i_{1}=I_{c}
J_{1}(I_{\omega}/I)$ \cite{Russer}. Here $J_{1}$ is the Bessel
function of the first kind.  As can be seen in Fig.  \ref{Iceff} the
temperature dependence of the amplitude $i_{1}$ strictly follows that
of $I_{c}$.  The small observed ratio $i_{1}/I_{c} \approx 0.06$
allows an estimate of the $a.c.$ excitation current $I_{\omega}
\approx 10 \, \mu$A at -20 dBm.  The treatment of the microwave
current as a non-perturbative probe is hence fully justified.  The
most striking observation is the non-monotonic temperature dependence
of both $i_{1/2}$ and $i_{1/3}$, in strong contrast with the
monotonous behavior of $i_{1}$.  As seen in Fig.  \ref{Iceff},
$i_{1/2}$ starts increasing at temperature of about 3 K, then reaches
the same amplitude as $i_{1}$ near 4 K, and finally slowly decreases
at higher temperatures.  Let us point out that the temperature at the
maximum is much larger than the Thouless temperature
$\epsilon_{c}/k_{B}$ = 0.28 K. A similar behavior is found on the
smaller $\frac{1}{3}$ step.  The persistence of the $\frac{1}{2}$ and
$\frac{1}{3}$ at high temperatures, $i.e.$ when the $d.c.$ Josephson
coupling is suppressed, is the most important result of this work.

The persistence of (integer) Shapiro-like steps at high temperatures
\cite{Volkov} is expected to occur in a four-terminal configuration
under current injection from a normal reservoir, but not in our
two-terminal geometry.  One can also discard the inductance or the
capacitance of the junction itself or the onset of chaos as a possible
origin of sub-harmonic steps \cite{Likharev-rsj}, as our
high-conductance junctions are strongly overdamped.  The sub-harmonic
steps must indeed be viewed as a manifestation of a non-sinusoidal
Josephson current-phase $I(\chi)$ relation.  An intrinsic
non-sinusoidal contribution to the $d.c.$ supercurrent is expected at
zero temperature but strongly suppressed as the temperature increases
\cite{Kulik}.  In a long junction at $k_{B}T>5 \, \epsilon_{c}$
($\approx 2$ K for our samples) the second harmonic is expected to
contribute to less than 2 $\%$ of the total current.  In contrast, the
sub-harmonic steps observed here not only appear as the temperature
increases, T $>$ 3 K as seen on Fig.  \ref{Iceff}, but they eventually
exceed the integer steps.  The intrinsic non-sinusoidal phase
dependence of the $d.c.$ Josephson current therefore cannot explain
the sub-harmonic steps.  The non-equilibrium $a.c.$ Josephson coupling
model \cite{Argaman} introduced in our discussion of the excess
current was shown to predict a finite $\sin 2\chi$ term even in
absence of $d.c.$ Josephson critical current.  Lehnert et al. 
\cite{Lehnert} reported such observation in InAs based junctions in
the clean limit.  Our results in the diffusive regime show strong
similarities with \cite{Lehnert} although the latter ones are
concerned with higher electron energy scale because of the smaller
electron transit time.

Let us now discuss the implications of sub-harmonic steps in terms of
coherent MAR processes.  The bias voltage involved here is so low that
the phase difference is quasi-static and low-energy Andreev electron
pairs are phase-coherent over the junction length.  This is the regime
where electrons close to the Fermi level experience coherent MAR. The
$\frac{1}{2}$ step is therefore associated with the coherent transfer
across the diffusive metal of double pairs.  Along the same way, the
$\frac{1}{3}$ step corresponds to the transfer of triple pairs.  In
the case of an interface transparency close to 1, this interpretation
holds provided the phase-breaking events can be neglected.  In this
respect, the characteristic time scale $2n \tau_{D}= 2n \times$ 28 ps
has to be smaller or at least comparable to the phase coherence time
$\tau_{\varphi} \simeq$ 160 ps at 4 K \cite{Roukes}.  At this
temperature, the condition is fulfilled for $n$ = 2 and 3, which is
consistent with our observations of $\frac{1}{2}$ and $\frac{1}{3}$
steps at low power.  Higher-order steps up to $n$ = 5 arise only at
high power when the inelastic damping is compensated by a larger
$a.c.$ current.

In summary we have investigated the low-voltage coherent transport in
diffusive S-N-S junctions with high barrier transparency.  We observed
a series of sub-harmonic Shapiro steps at high temperature and low
voltage $eV < \epsilon_{c} < k_{B}T$ where ordinary Josephson coupling
is known to vanish.  Presently there is no satisfactory description of
the microscopic mechanism which is very likely to involve coherent
transfer of multiple charges.

We acknowledge fruitful discussions with Y. Blanter, M. B\"{u}ttiker,
J. C. Cuevas, K. Lehnert, A. Levy-Yeyati, F. Wilhelm, A. Zaikin and
within the TMR program "Dynamics of superconducting nanocircuits".  We
thank F. Hekking for pointing out the relevance of coherent MAR.

\end{document}